\documentclass{article}
\usepackage{spconf,amsmath,graphicx}

\usepackage{array}
\usepackage{amsfonts}
\usepackage{tikz}
\usepackage{comment}
\usepackage{color}
\usepackage{booktabs}
\usepackage{xspace}
\usepackage{amssymb}
\usepackage{amsfonts}
\usepackage{mathrsfs}
\usepackage{subfigure}
\usepackage{multirow}
\usepackage{graphicx}
\usepackage{enumitem}
\usepackage{mathtools}
\usepackage{xcolor}
\usepackage{url}
\usepackage{hyperref}
\usepackage{cite}
\usepackage{arydshln}
\usepackage{pgfplots}
\pgfplotsset{width=7cm, compat=1.17}
\usepackage{algorithm}
\usepackage{algpseudocode}
\def\presec{\vspace{-1em}}
\def\postsec{\vspace{-0.5em}}

\newcommand{\argmax}{\operatorname{argmax}}

\title{Fast and parallel decoding for transducer}
\name{Wei Kang\textsuperscript{1}, Liyong Guo\textsuperscript{1}, Fangjun Kuang\textsuperscript{1}, Long Lin\textsuperscript{1}}
\myname{Mingshuang Luo\textsuperscript{1}, Zengwei Yao\textsuperscript{1}, Xiaoyu Yang\textsuperscript{1}, Piotr Żelasko\textsuperscript{2}, Daniel Povey\textsuperscript{1}}

\address{\textsuperscript{1} Xiaomi Corp., Beijing, China \:\:\: \textsuperscript{2}Meaning.Team Inc, USA \\
\footnotesize{\texttt{\{kangwei1, dpovey\}@xiaomi.com, pzelasko@meaning.team}}}

\begin{document}
%\ninept
%
\maketitle
\begin{abstract}

The transducer architecture is becoming increasingly popular in the
field of speech recognition, because it is naturally streaming as well as high in accuracy. One of the drawbacks
of transducer is that it is difficult to decode in a fast and parallel way due to
an unconstrained number of symbols that can be emitted per time step.

In this work, we introduce a constrained version of transducer loss to learn strictly monotonic alignments between the sequences; we also improve the standard greedy search and beam search algorithms
by limiting the number of symbols that can be emitted per time step in transducer
decoding, making it more efficient to decode in parallel with batches. Furthermore, we propose an finite state automaton-based (FSA) parallel beam search algorithm that can run with graphs on GPU efficiently. The experiment results show that we achieve slight word error rate (WER) improvement as well as significant speedup in decoding. Our work is open-sourced and publicly available\footnote{https://github.com/k2-fsa/icefall}.

\end{abstract}
\begin{keywords}
speech recognition, transducer, end-to-end, beam search, parallel decoding
\end{keywords}

\presec
\section{Introduction}
\postsec
\label{sec:introduction}

The transducer architecture \cite{graves2012sequence} has been growing in popularity
in the field of automatic speech recognition (ASR), especially for deployed
real-time ASR systems \cite{sainath2020streaming, he2019streaming} because it
support streaming naturally while achieving high accuracy

Unlike decoding with CTC models where at most one symbol can be emitted per
time step, the number of symbols emitted per time step is unconstrained in
transducer. Therefore, it is difficult to perform parallel decoding with transducer,
because the number of sequential operations is hard to bound in advance.

Various efforts have been made to accelerate transducer decoding.
In \cite{he2019streaming}, caching techniques are used to avoid redundant
computation in the prediction network for identical prediction histories.
Pruning is used in \cite{jain2019rnn} to reduce the number of active hypotheses
during the search to make computation more efficient.
Another way to reduce the number of active hypotheses is proposed in
\cite{prabhavalkar2021less} by using a prediction network with limited label
context to merge hypotheses with identical prediction histories.

Different from the above-mentioned works that still allow unlimited number of
symbols per time step, we limit the number of symbols that can be emitted per
time step to 1. The most similar work to ours is \cite{kim2020accelerating} and \cite{tripathi2019monotonic}, both of which constrain the hypotheses expanded at each decoding step to one. However,
there are two differences between our work and \cite{kim2020accelerating}.
First, we also pose the constraint on model training so that it has the same
behavior in training and decoding. Second, \cite{kim2020accelerating} uses two
transitions in the transducer lattice during decoding, first going upward and
then rightward, while there is only one diagonal transition in our work, which
can further save computation and is easier to decode in batch mode. In addition, we use a different transducer architecture from \cite{tripathi2019monotonic} to learn monotonic alignments between the sequence, which shows more promising results. 

% \begin{comment}
The main contributions of this paper are:
\begin{itemize}[itemsep=1.8pt,topsep=1.8pt,parsep=0pt]
  \item We accelerate the transducer decoding by limiting the number of symbols emitted per time step to one.
  \item A constrained transducer is proposed to improve the performance of the one-symbol-per-frame decoding.
  \item We implement an FSA-based parallel beam search algorithm that can run with graphs on GPU efficiently.
\end{itemize}
% \end{comment}

\begin{comment}
The rest of this paper is organized as follows.
Section~\ref{sec:algo} analyzes transducer decoding algorithms from different aspects
and describes the proposed decoding algorithm.
Section~\ref{sec:constrained-rnnt} and Section~\ref{sec:pruned-rnnt}describes how we train the models to match our proposed decoding method. The experiment setup and results are given in Section~\ref{sec:experiments}. Finally, we conclude
the paper in Section~\ref{sec:conclusions}.
\end{comment}
\presec
\section{Transducer decoding algorithms}
\postsec
\label{sec:algo}

We will first analyze the state-space and transitions of transducer decoding algorithms in detail, during which we show how we simplify the decoding process step by step. Then we will describe the implementation details of our proposed FSA-based beam search algorithm.

\presec
\subsection{Traditional Transducer}
\postsec

With the vanilla, fully-recurrent transducer, the \textbf{state-space} consists of
pairs $(\mathbf{y}, t)$ where $\mathbf{y}$ is a sequence of symbols (excluding
$\varnothing$) and $t$ is an integer frame index.  If there are $T$ frames
numbered $0, 1, \ldots, T{-}1$, then $((), 0)$ is initial (i.e. empty-sequence,
0) and states $(\mathbf{y}, T)$ are final.

The \textbf{transitions} are as follows: for symbols $a \neq \varnothing$ and
for $0 \leq t < T$, there are transitions $(\mathbf{y}, t) \rightarrow (\mathbf{y} + a, t)$
with label $a$ and probability $P(a | \mathbf{y}, t)$.
Also $(\mathbf{y}, t)$ has a transition to $(\mathbf{y}, t{+}1)$ with probability
$P(\varnothing | \mathbf{y}, t)$.  There are no cycles in this graph because
transitions are always either to larger $t$ or to (same $t$, longer $\mathbf{y}$).

\presec
\subsection{Stateless Transducer}
\postsec

In the stateless transducer as in~\cite{ghodsi2020rnntstateless} the decoder network requires only a finite
context, for instance two symbols.
Therefore, the state-space can be reduced to only the most recent symbols, e.g. two
symbols of ${\mathbf y}$; the initial state can be taken to be $((\varnothing, \varnothing), 0)$,
and when a transition $c \neq \varnothing$ is made from state $((a, b), t)$, $((b, c), t)$\footnote{Our FSA-based decoding uses this reduced search-space, but our
non-FSA-based beam search algorithm still uses the full sequence $({\mathbf y}, t)$
as the state space (so we can re-use the previous beam-search code).} is reached.
% Our FSA-based decoding uses this reduced search-space, but our
% non-FSA-based beam search algorithm still uses the full sequence $({\mathbf y}, t)$
% as the state space (so we can re-use the previous beam-search code).

\presec
\subsection{Max-symbols decoding}
\postsec

Here, we describe decoding methods where we limit the maximum number
of symbols $S$ that can be emitted per frame, to 1, 2 or 3.  The state-space
is extended by a number $0 \leq n < S$ saying how many symbols we have
already emitted on this frame, so a state would be of the form $({\mathbf y}, t, n)$.
Transitions with a blank label ($\varnothing$) are: $({\mathbf y}, t, n) \rightarrow ({\mathbf y}, t{+}1, 0)$.
For $n \leq S{-}1$, transitions with label $a \neq \varnothing$ are:
$({\mathbf y}, t, n) \rightarrow ({\mathbf y}{+}a, t, n{+}1)$.
For $n = S{-}1$, transitions with labels $a \neq \varnothing$ are:
$({\mathbf y}, t, S{-}1) \rightarrow ({\mathbf y}{+}a, t{+1}, 0)$.
This is equivalent to assuming the probability of blank is always $1$ after emitting $S$ symbols on a given frame. 
%Our approach is thus a little different from~\cite{kim2020accelerating}, where they actually compute the last blank probability (we do it our way for efficiency).  So in their version of the algorithm, $0 \leq n \leq S$ and non-blank transitions from states $({\mathbf y}, t, S)$ are simply disallowed.

The expansion of the state space in max-symbols decoding does not in itself affect the decoding result:
for sufficiently large $n$, the sequence ${\mathbf y}$ of the best path will still
be the same, and the ${\mathbf y}$ with the greatest total probability will also
be the same, as it is in conventional transducer decoding.  This is because there is a one-to-one
mapping between paths in the conventional and max-symbols decoding algorithm
(for large $n$).

\presec
\subsection{FSA-based decoding}
\postsec

For simplicity, our FSA decoding algorithm assumes:
\begin{itemize}[itemsep=1.8pt,topsep=1.8pt,parsep=0pt]
  \item We are using stateless transducer where the decoder consumes a fairly
     small number of input symbols\footnote{This makes it possible to encode the history sequence into a single integer.}.
   \item We are doing {\em max-symbols} decoding with $S = 1$.  This
     is quite similar to hybrid or CTC decoding, since all transitions are to the next frame.
\end{itemize}
We further extend the state-space to enable decoding with
graphs (like conventional hybrid decoding, except with no hidden Markov model topology).
Taking the history-length of stateless transducer to be 2, the states are of the
form $((a, b), t, s)$ where $a$ and $b$ are symbols (possibly $\varnothing$ if we
are near the start of the utterance), $t \geq 0$ is the frame-index, and $s$ is the decoding-graph
state.  Thus, our decoding algorithm implements graph composition.  For each
arc $s \rightarrow r$ in the decoding graph with label $c \neq \varnothing$, and
probability $q$, there exists a transition in the lattice $((a, b), t, s) \rightarrow ((b, c), t{+}1, r)$ with
label $c$ and probability $q P(c | (a, b), t)$.
In addition, we have blank transitions $((a, b), t, s) \rightarrow ((a, b), t{+}1, s)$
with probability $P(\varnothing | (a, b), t)$ and label $\varnothing = 0$.
The graph is assumed to be epsilon-free.

Our decoding algorithm is implemented in k2~\footnote{https://github.com/k2-fsa/k2} using ragged tensor data structures, which enables rapid processing of  irregular-sized objects in parallel on the GPU.

\vspace{-2em}
\subsection{Beam search}
\postsec
It is hard to compactly describe the search strategy
of the algorithm in~\cite{graves2012sequence}, but the goal generally seems to be to keep
no more than the $N$ best paths active; and the frames are processed one by one.  The order of processing within a frame is ``best-first''.
This is not consistent with the goal of summing probabilities, because
we may have already processed a sequence before its prefix; also, the algorithm
generates duplicates and is not very explicit about how to deal with this.
People have implemented slightly different versions of this algorithm,
e.g.~ESPNet~\cite{watanabe2018espnet} dispenses with the ``prefix search''
part of the algorithm and instead implements de-duplication of hypotheses.

For our proposed \textbf{FSA-based decoding}, all transitions are from frame $t$ to $t{+}1$
because we use {\em max-symbols}$=1$.
There are 3 constraints: a log-probability beam, a {\em max-states} constraint
(that limits the number of tuples $((a, b), t, s)$ for a given $t$), and a
{\em max-contexts} constraint that limits the number of symbol contexts
like $(a,b)$ that are active on a given $t$. On each frame we first
do propagation to the next frame without pruning.  We then apply
the {\em max-states} and {\em beam} constraints in one pruning operation;
and then apply the {\em max-contexts} constraint in a second pruning operation.

The pseudocode of our \textbf{FSA-based decoding} is given in Algorithm ~\ref{alg:fast_beam_search} \footnote{We call it fast\_beam\_search in icefall (https://github.com/k2-fsa/icefall).}, it can decode many streams in parallel. The output of our algorithm is a lattice (i.e. an FSA), we can find either the best path or the
$\varnothing$-free label sequence ${\mathbf y}$ with the highest probability, using standard FSA operations, after
generating the lattice.  The label sequence with the highest probability is found by
\begin{itemize}[itemsep=1.8pt,topsep=1.8pt,parsep=0pt]
  \item Generating $n$-best paths from the lattice using an easy-to-parallelize randomized algorithm.
  \item Finding the unique paths ${\bf y}$ from the $n$-best paths by removing $\varnothing$ from the label sequences.
  \item Generating FSAs from the unique paths, composing these with the lattice, and computing the total
      probability of the results.
\end{itemize}

\begin{algorithm}[ht]
\footnotesize
\caption{FSA-based Parallel Beam Search}
\label{alg:fast_beam_search}
\begin{algorithmic}[1]
\Require 
The decoding graphs for each sequence $fsas$;

The pruning parameters $beam$, $max\_contexts$, $max\_states$;

The instance of transducer model $M$;

The input acoustic features $feats$;
\State $Streams \gets []$
\For {$i \gets 0$ to $num\_sequences$};
\State $Streams[i] \gets Init(fsas[i], beam, max\_contexts, max\_states)$;
\EndFor
\For {$t \gets 0$ to $T$};
\State $Enc_t \gets M.encoder(feats[t])$ 
\State $shape~\footnotemark, contexts \gets GetContexts(Streams)$
\State $Dec_{out} \gets M.decoder(contexts)$
\State $Enc_{out} \gets IndexSelect(Enc_t, shape.RowIds(1))$
\State $log\_probs \gets M.joiner(Enc_{out}, Dec_{out})$
\State $Streams.ExpandArcs(log\_probs)$
\State $Streams.Prune()$
\EndFor
\State $Streams.TerminateAndFlushToStreams()$
\State $lattice \gets Streams.FormatOutput()$
\State \textbf{return} $lattice$
\end{algorithmic}
\end{algorithm}
\footnotetext{This is a RaggedShape in k2 (https://github.com/k2-fsa/k2)}

\presec
\section{Constrained Transducer Training}\label{sec:constrained-rnnt}
\postsec

Since we find that decoding with {\em max-symbols}${=}1$ works well, so that in the decoding
algorithm, consuming a non-$\varnothing$ symbol takes us from frame $t$ to $t+1$, it is natural to try incorporating this rule in training as well. 

\presec
\subsection{Modified Transducer}
\postsec

We first try training the transducer by introducing a diagonal transition in the transducer lattice, just like the way that \cite{tripathi2019monotonic} did, where emitting a non-blank symbol takes you to the next frame. We call this ``modified transducer''.  If the regular core transducer recursion is:
\begin{align}
\alpha(t,u) = \mathrm{log\_add}&\left( \alpha(t-1, u)+ \varnothing(t-1, u) \right.,\nonumber\\
                               &\left. \alpha(t, u-1) +  y(t, u-1) \right) \label{eq:standard-forward}
)
\end{align}
with the final data-likelihood being $\alpha(T{-}1,U) + \varnothing(T{-}1, U)$,
the modified transducer is:
\begin{align}
\alpha(t,u) = \mathrm{log\_add}&\left( \alpha(t-1, u)+ \varnothing(t-1, u) \right.,\nonumber\\
                               &\left. \alpha(t-1, u-1) +  y(t-1, u-1) \right) \label{eq:modified-forward}
)
\end{align}
with the final data-likelihood being $\alpha(T,U)$.

\presec
\subsection{Constrained transducer}
\postsec

We also propose a new architecture called constrained transducer, which is like modified transducer in that you have to go to next frame when you emit a non-blank system, but this is done by "forcing" you to take the blank transition from the \textbf{next} context on the \textbf{current} frame, e.g. if we emit c given "a b" context, we are forced to emit "blank" given "b c" context on the current frame. The core recursion is:
\begin{align}
& \alpha(t,u) = \mathrm{log\_add}\left( \alpha(t-1, u)+ \varnothing(t-1, u) \right.,\nonumber\\
                               &\left. \alpha(t-1, u-1) +  y(t-1, u-1) + \varnothing(t-1,u) \right) \label{eq:constrained-forward}
)
\end{align}
with the final data-likelihood being $\alpha(T,U)$.

\presec
\section{Pruned RNN-T training and lm-scale} \label{sec:pruned-rnnt}
\postsec

We proposed the pruned RNN-T training in our previous work~\cite{fjkuang2022prunedrnnt}.
This is a more efficient way of evaluating the RNN-T recursion, by using a ``trivial'' joiner
network to quickly evaluate the recursion and figure out which $(t, u)$ pairs are important,
then only evaluating the full joiner on a subset of symbols.  For the purpose of this work, the details on pruning are emitted. What is important to know is that
as part of the pruned RNN-T training, we regularize the loss function with
a {\em trivial joiner}, which simply adds logprobs derived from the
encoder with logprobs from the decoder; and we interpolate the trivial-joiner logprobs with
$\alpha^{\mathrm{lm}}$, or {\em lm-scale}, times logprobs derived from the decoder alone.
This essentially means that we are including a language-model probability (predicting the
tokens or blanks) in the log-probs used in the simple-joiner recursion.
This {\em lm-scale} term forces the decoder log-probs used in the trivial joiner to be close to the probabilities of
a ``real'' language model (on the vocabulary including $\varnothing$).  For reasons
that are currently unclear to us, this appears to affect the model in some way that
makes decoding with {\em max-symbols}$=1$ work better, and also slightly improves
WERs.  The trivial joiner that is being affected by this regularization
is not used in decoding, so it must be an indirect effect that acts by
changing the encoder output in some way.
\presec
\section{Experiments}
\label{sec:experiments}

\vspace{-0.5em}
\subsection{Dataset and Setup}
\postsec
We conduct all our experiments on the popularly used Librispeech corpus~\cite{librispeech2015}. Lhotse~\cite{Zelasko_Lhotse_a_speech_2021} is used for data preparation. The acoustic feature is 80-channel Fbank extracted from a 25ms window with a stride of 10ms. We also use spec-augmentation~\cite{park2019specaug} and noise augmentation (by mixing MUSAN~\cite{musan2015}) to improve generalization. Furthermore, speed
perturbation~\cite{ko2015audio} with factors 0.9 and 1.1 is used to triple the training set.

Our encoder model is a re-worked version of Conformer~\cite{anmol2020conformer}. It has 12 encoder layers, each of which contains 8 self-attention~\cite{vaswani2017attention}
heads with attention-dim 512.
The subsampling factor in the convolution module is 4 and the feed forward
dimension is 2048. The decoder model is a stateless network~\cite{ghodsi2020rnntstateless},
consisting of an embedding layer with embedding dimension 512 followed by a 1-D
convolution layer with kernel size 2. The outputs of the model are 500 sentence
pieces~\cite{kudo-richardson-2018-sentencepiece} with byte pair encoding (BPE). All of the models are trained with pruned RNN-T loss~\cite{fjkuang2022prunedrnnt}.
%

% \begin{figure*}[t]
% \includegraphics[width=\textwidth, trim={0 1cm 0 1cm},clip]{lattice.pdf}
% \vspace*{-10mm}
% \caption{Part of the decoded lattice for an utterance generated by FSA-based beam search.
% It is interesting to note that tokens can appear in successive frames.
% For example, the token 45
% can appear on the following transitions: (1) From the state 73 to the state 74;
% (2) from the state 75 to the state 76; (3) from the state 78 to the state 79.
% }
% \label{fig:lattice}
% \vspace*{-5mm}
% \end{figure*}

\presec
\subsection{ASR accuracy}
\postsec

\begin{table}[ht]
\vspace{-1em}
\footnotesize
\begin{center}
\caption{\small{The max-symbols decoding results (WER) of different architecture of transducers \textbf{(without lm\_scale)}.}} \label{tbl:all_rnnt}
\begin{tabular}{lp{0.5cm}p{0.5cm}p{0.5cm}p{0.5cm}p{0.5cm}p{0.5cm}p{0.5cm}}
\hline
Setup                          &  Max        & \multicolumn{2}{c}{Greedy}  & \multicolumn{2}{c}{Beam}      & \multicolumn{2}{c}{FSA based}   \\
                               &  Symbol     & \multicolumn{2}{c}{Search}  & \multicolumn{2}{c}{Search}    & \multicolumn{2}{c}{Beam Search} \\
                               &             & clean& other              & clean    & other                & clean       & other             \\ \hline
\multirow{3}{*}{Regular}       & 1           & 2.72 & 6.33               & 2.67     & 6.21                 & 2.70        & 6.21              \\
                               & 2           & 2.68 & 6.28               & /        &  /                   & /           & /                 \\
                               & $\infty$    & 2.68 & 6.28               & 2.65     & 6.18                 & /           & /                 \\ \hline
\multirow{3}{*}{Constrained}   & 1           & 2.76 & 6.46               & 2.74     & 6.35                 & 2.74        & 6.33              \\
                               & 2           & 2.76 & 6.46               & /        &  /                   & /           & /                 \\
                               & $\infty$    & 2.76 & 6.46               & 2.74     & 6.32                 & /           & /                 \\ \hline
\multirow{3}{*}{Modified}      & 1           & 3.68 & 9.43               & 3.54     & 9.07                 &  3.57       &  9.03             \\
                               & 2           & 70.9 & 70.7               &  /       &  /                   & /           &  /                \\
                               & $\infty$    & 1179 & 1089               &  -       &   -                  & /           &  /                \\ \hline
\end{tabular}
\end{center}
\vspace*{-5mm}
\end{table}

\begin{table}[ht]
\vspace{-1em}
\footnotesize
\begin{center}
\caption{\small{The max-symbols decoding results (WER) of different architecture of transducers \textbf{(with lm\_scale=0.25)}.}}\label{tbl:lm_scale_rnnt}
\begin{tabular}{lp{0.5cm}p{0.5cm}p{0.5cm}p{0.5cm}p{0.5cm}p{0.5cm}p{0.5cm}}
\hline
Setup                          &  Max        & \multicolumn{2}{c}{Greedy}  & \multicolumn{2}{c}{Beam}      & \multicolumn{2}{c}{FSA based}   \\
                               &  Symbol     & \multicolumn{2}{c}{Search}  & \multicolumn{2}{c}{Search}    & \multicolumn{2}{c}{Beam Search} \\
                               &             & clean& other              & clean    & other                & clean       & other             \\ \hline
\multirow{3}{*}{Regular}       & 1           & 2.66 & 6.25               & 2.62     &  6.08                & 2.67        & 6.20              \\
                               & 2           & 2.66 & 6.25               &  /       &  /                   & /           & /                 \\
                               & $\infty$    & 2.66 & 6.25               & 2.62     &  6.10                & /           & /                 \\ \hline
\multirow{3}{*}{Constrained}   & 1           & 2.61 & 6.45               & 2.6      & 6.39                 & 2.64       & 6.39               \\
                               & 2           & 2.61 & 6.45               & /        & /                    & /           & /                 \\
                               & $\infty$    & 2.61 & 6.45               & 2.6      & 6.39                 & /           & /                 \\ \hline
\multirow{3}{*}{Modified}      & 1           & 2.62 & 6.41               & 2.59     & 6.33                 & 2.64        & 6.32              \\
                               & 2           & 6.35 & 10.6               & /        &  /                   & /           &  /                \\
                               & $\infty$    & 108  & 114                & -        &  -                   & /           &  /                \\ \hline
\end{tabular}
\end{center}
\vspace*{-5mm}
\end{table}

Table~\ref{tbl:all_rnnt} shows the max-symbols decoding results of three types of transducers.
For regular transducer, the WERs increase when the number of symbols per frame is constrained during decoding,
while there are no degradations in WERs for constrained transducer. We also find that the modified transducer produces much worse results than the other two, and there will be a lot of insertion errors when we increase the number of symbols that can be emitted per frame. It seems that during the training, blank probability is important when transiting to the next frame given a non-blank symbol.

By comparing Table~\ref{tbl:lm_scale_rnnt} with Table~\ref{tbl:all_rnnt}, it shows that the $lm\_scale$ in the pruned RNN-T loss not only makes decoding with {\em max-symbols}$=1$ work better, but also helps to improve the performance.

\begin{table}[ht]
\vspace{-1em}
\footnotesize
\begin{center}
\caption{\small{The effects of how duplicated alignments of the same hypotheses are handled.}}\label{tbl:merge_operation}
\begin{tabular}{lccccc}
\hline
Decoding                                 &  Max        & \multicolumn{4}{c}{Merge OP}      \\
Method                                   &  Symbol     & \multicolumn{2}{c}{Max}        & \multicolumn{2}{c}{LogAdd}           \\
                                         &             & clean  & other                   & clean    & other                   \\ \hline
Beam                                     & 1           & 2.63   & 6.37                    & 2.60     & 6.39                    \\
Search                                   & $\infty$    & 2.64   & 6.37                    & 2.60     & 6.39                    \\ \hline
\parbox[c]{2cm}{FSA based \\beam search} & 1           & 2.64   & 6.39                    & 2.66     & 6.42                    \\ \hline
\end{tabular}
\end{center}
\vspace*{-5mm}
\end{table}

\begin{table}[ht]
\vspace{-1em}
\footnotesize
\begin{center}
\caption{\small{The effects of length normalization to WERs.}}\label{tbl:length_norm}
\begin{tabular}{lccccc}
\hline
Decoding                                 &  Max        & \multicolumn{4}{c}{Length Norm}      \\
Method                                   &  Symbol     & \multicolumn{2}{c}{(Yes)}        & \multicolumn{2}{c}{(No)}           \\
                                         &             & clean  & other                   & clean    & other                   \\ \hline
Beam                                     & 1           & 2.60   & 6.39                    & 2.60     & 6.40                    \\
Search                                   & $\infty$    & 2.60   & 6.39                    & 2.60     & 6.38                    \\ \hline
\end{tabular}
\end{center}
\vspace*{-5mm}
\end{table}

The stated goal of the beam search algorithm presented in~\cite{graves2012sequence} seems to be to find the ${\mathbf y}$ with the largest
length-normalized probability $\argmax_{\mathbf y} p(\mathbf{y})$. So we also investigated the effects of length normalization and the way how we handle duplicate alignments of
the same hypotheses. Table~\ref{tbl:merge_operation} and ~\ref{tbl:length_norm} show that both of them have little impact on the WER.

% Figure~\ref{fig:lattice} shows part of the decoded lattice for an utterance
% generated by FSA-based beam search. It is worth noting that tokens can appear
% in successive frames in the lattice. For instance, the token 45 can appear in
% 3 successive frames.

\presec
\subsection{Decoding speed}
\postsec

\begin{table}[ht]
\vspace{-1em}
\footnotesize
\begin{center}
\caption{\small{The RTF of different decoding methods.}}\label{tbl:rtf}
\begin{tabular}{lccc}
\hline
Decoding                                 & Max      & \multicolumn{2}{c}{RTF}     \\
Method                                   & Symbol   & batched(No)  & batched(Yes) \\ \hline
\multirow{2}{*}{Greedy Search}           & 1        &  0.011   & 0.0009   \\
                                         & $\infty$ &  0.011   & 0.0078   \\ \hline
\multirow{2}{*}{beam search}             & 1        &  0.03    & 0.0068   \\
                                         & $\infty$ &  0.126   & 0.125    \\ \hline
\parbox[c]{2cm}{FSA based \\beam search} & 1        &  0.05    & 0.002    \\ \hline
\end{tabular}
\end{center}
\vspace*{-5mm}
\end{table}

Table~\ref{tbl:rtf} illustrates the real-time factor (RTF) for the test-clean and test-other datasets using different decoding methods. We conduct the benchmark on an NVIDIA V100 GPU with 32 GB RAM, when running in batches, dynamic batch size is used to make full use of the memory. For greedy search and beam search that are hard to support parallel decoding with max-symbol$=\infty$, only the encoder network runs in batches when batched equals "Yes". The results show that there is a clear speed advantage in decoding by limiting the number of symbols to 1, especially when running in parallel. It also shows that our proposed FSA-based beam search is 3.4 times faster than the standard beam search in ~\cite{graves2012sequence}.

\presec
\subsection{Decoding with FSA}
\postsec

\begin{table}[ht]
\vspace{-1.5em}
\footnotesize
\begin{center}
\caption{\small{The WERs and RTF of FSA based Beam Search with trivial graph and LG graph.}}\label{tbl:decoding_lg}
\begin{tabular}{lcccc}
\hline
Decoding                        & Merge OP   & \multicolumn{2}{c}{WERs}  & RTF       \\
Graph                           &            & clean    & other            &         \\ \hline
\multirow{2}{*}{Trivial Graph}  &  Max       &  2.64    & 6.39             & 0.0024  \\
                                &  LogAdd    &  2.66    & 6.42             & 0.0112  \\ \hline 
\multirow{2}{*}{LG Graph}       &  Max       &  2.84    & 6.37             & 0.0025  \\ 
                                &  LogAdd    &  2.86    & 6.36             & 0.004   \\ \hline
\end{tabular}
\end{center}
\vspace*{-5mm}
\end{table}

The FSA-based beam search results in table~\ref{tbl:all_rnnt}~\ref{tbl:lm_scale_rnnt}~\ref{tbl:merge_operation}~\ref{tbl:rtf} are all decoded with a “trivial graph” that has only one state. Table~\ref{tbl:decoding_lg} gives the WERs and RTF for FSA-based beam search with a general graph. The G in LG graph is a 3-gram LM trained on Librispeech text corpus, which has around 180MB in arpa text format. From the table, we can see that using a larger FSA graph won't affect the RTF too much, which shows  that our FSA-based beam search algorithm also works efficiently on general FSA graphs. As for the performance degradation on LG graph, we find that there are some words not present in the released lexicon, making the results on LG graph worse than the trivial graph. As for the larger RTF on trivial graph using log\_add comparing with LG graph, it is because the decoding lattice generated with trivial graph is larger than that generated with LG graph, which makes it take more time to find the label sequence with the highest probability .  

\presec
\section{Conclusions}
\postsec
\label{sec:conclusions}

In this paper, we improve the transducer decoding algorithms by limiting the number of symbols that can be emitted per time step to one, making it possible to decode in parallel using batches. We also propose a constrained transducer architecture and $lm\_scale$ training in our pruned RNN-T loss that make the decoding with max-symbol$=1$ work better. What's more, we have implemented a fast and highly-parallel FSA-based decoding algorithm for transducer that generates lattices; it gives slightly better accuracy and up to 3.4 times faster performance than conventional transducer decoding.

\newpage

\bibliographystyle{IEEEbib}
\bibliography{strings,refs}

\end{document}